\documentclass[superscriptaddress,aps,prl,floats,twocolumn,twoside,floatfix]{revtex4}

\usepackage{epsfig}
\usepackage{amsmath}
\usepackage{amssymb}

\renewcommand{\epsilon}{\varepsilon}

\begin{document}

\title{Shear-thickening and entropy-driven reentrance}

\author{Mauro Sellitto}
\affiliation{Institute for Scientific Interchange, Viale
  S. Severo 65, 10133 Torino, Italy}

\author{Jorge Kurchan}
\affiliation{PMMH-ESPCI, CNRS UMR 7636, 10 rue Vauquelin, 75005
  Paris, France}

\date{\today}

\begin{abstract}
  We discuss a generic mechanism for shear-thickening analogous to
  entropy-driven phase reentrance.  We implement it in the context of
  non-relaxational mean-field glassy systems: although very simple,
  the microscopic models we study present a dynamical phase diagram
  with second and first order stirring-induced jamming transitions
  leading to intermittency, metastability and phase coexistence as
  seen in some experiments. The jammed state is fragile with respect
  to change in the stirring direction.  Our approach provides a direct
  derivation of a Mode-Coupling theory of shear-thickening.
\end{abstract}

\maketitle

When liquids are subjected to strong stirring, in general their
viscosity decreases, a phenomenon known as `shear thinning'. The
opposite (`shear thickening') behavior when stirring leads to
increased jamming is rather exceptional and intriguing~\cite{larson}.
That shear-thinning should be generic is quite easy to understand by
considering the structure of any system having a large viscosity and
long relaxation time scales.  From the point of view of the
phase-space energy landscape, the long relaxation times are the
consequence of directions that are either almost flat, or contain
barriers that can only just be crossed.  When the system is stirred by
non-conservative forces, displacements are easily induced along these
directions, and this has the effect of speeding the relaxation.  From
the real space point of view, slow relaxations are linked to extended
dynamically correlated spatial regions~\cite{heter}, and stirring
tends to break off these domains, thus making the system more fluid.
It is thus no surprise that just about any system (and any model) will
naturally exhibit shear-thinning.

For shear-thickening instead, several explanations have been
attempted, and it is at present not clear whether a universal one will
apply for all possible systems.  There are indications that jamming in
particulate suspensions is related to increased
disorder~\cite{hoffman}, and in some cases to the formation of
clusters of particles in lubrication contact~\cite{brady,ball}.  To
the extent that at large densities the strongly jammed state has the
appearance of an amorphous, glassy solid, shear-thickening may be
thought of as a consequence of an underlying glass transition induced
by stirring~\cite{holmes}. For such collective behavior one can
attempt a theory with less focus on the details but founded on notions
that are thought to be generic of glasses: this has suggested, for
example, the casting of the problem in a Mode Coupling
format~\cite{holmes}.

In this Letter we attempt a microscopic setting in which glassiness
and shear-thickening emerge naturally and are simultaneously
understood.  The basic idea is to exploit the analogy between
entropy-driven transitions in which systems freeze upon heating and
those in which they jam under the action of stirring.  To obtain
concrete results, we discuss this idea within the `Random First Order'
scenario for the glass transition, although it is not restricted to
it.  In its simplest version, the Random First Order scenario applies
to models of fully connected degrees of freedom and a complicated set
of interactions with or without quenched
disorder~\cite{wolynes,reviews}.  It allows for a unified description
of the (fragile) glass transition, both from the dynamical and from
phase-space landscape points of view. The approach contains as a
special, high temperature case the Mode Coupling Theory~\cite{Gotze},
while in the low temperature regime it provides a theory for
aging~\cite{reviews}.
As is characteristic of this approach, it has the satisfactory feature
that many different aspects of the collective behavior follow without
further assumptions, and the weakness that spatial features are not
(for the moment) fully incorporated.

{\bf Introducing stirring.} Shear-thinning appears naturally if one
considers the action of `stirring' terms capable of generating
permanent currents, i.e. forces that do not derive from a (global,
time-independent) potential.  A useful, though approximate, way to see
the effect of random stirring is the following: consider a system with
coordinates $x_i$ evolving according to some form of dynamics
(Langevin, Monte Carlo, molecular dynamics) in contact with a heat
bath at temperature $T$ and under the action of a potential and of
stirring forces $f_i^{\scriptscriptstyle \rm stir}$ acting on the
$i^{\scriptscriptstyle \rm th}$ degree of freedom.  Stirring forces
are by definition nonconservative, suppose (although this is
inessential) they are linear: $f_i^{\scriptscriptstyle \rm stir}=
J^{\scriptscriptstyle \rm as}_{ij}x_j$. If we make the simplifying
though rather crude assumption that the $J^{\scriptscriptstyle \rm
  as}_{ij}$ are long range, randomly distributed (so
$J^{\scriptscriptstyle \rm as}_{ij}$ is asymmetric) and uncorrelated,
one can easily show~\cite{sompo,Cukupe1} that on average
$f^{\scriptscriptstyle \rm stir}_i = \rho_i(t)$ where $\rho_i(t)$ are
Gaussian noises with correlations $\langle \rho_i(t)\rho_j(t')\rangle
=\delta_{ij} C(t,t')$ where $C(t,t')= \sum_k \langle x_k(t)x_k(t')
\rangle/N$ is the two-time autocorrelation function.  The stirring
thus provides a random noise unmatched by a friction term: this can be
seen as a coupling to an infinite temperature (self-consistent)
bath~\cite{note1}. Just like in any stirring situation, if the system
for some reason does not flow, the noise $\rho_i$ becomes
time-independent and hence does no work.

The generic situation when the (e.g. Monte Carlo) dynamics is
perturbed by a nonconservative force is that the structural $\alpha
$-relaxation time becomes shorter -- a shear-thinning effect -- and
that a two-temperature regime emerges~\cite{Babeku} even in the
supercooled liquid phase. This scenario has been tested in realistic
systems~\cite{Babe} and the agreement is impressive.  In contrast to
the case of shear-thinning, there has as yet been no way to introduce
or understand shear-thickening in these terms, and a phenomenological
construction with a Mode-Coupling flavor has to be introduced in a
somewhat {\em ad hoc} manner, with no underlying microscopic
model~\cite{holmes}.

{\bf A phase reentrance mechanism.} Let us review briefly a
microscopic mechanism~\cite{ScSh} for freezing induced by heating
(inverse freezing).  Suppose one has an ensemble of molecules
(e.g. polymers) that have a low temperature (`folded') state in which
they are mutually weakly interacting, and a higher temperature
(`unfolded') state which is favored entropically and in which they
interact strongly with each other. As temperature is increased, each
polymer unfolds and reaches out to the other polymers, the resulting
entangling thus may lead to a glass transition. A further increase of
temperature will eventually lead back to a liquid phase.

In order to obtain a minimal model of a liquid that upon heating is
driven by entropy into a glass, one can consider~\cite{ScSh} spins
taking values $0, \pm 1$, and a Hamiltonian consisting of a term
$\propto \sum_i s_i^2$ favouring the `folded' configurations $s_i=0$,
and an interaction term $\sum_{ij} J_{ij} s_i s_j$ that is active when
the spins are in the `unfolded' states $s_i=\pm 1$:
\begin{equation}
  H = - 2\sum_{ij} J_{ij} s_i s_j + D \sum_i s^2_i 
  \label{H}
\end{equation}
The entropic favouring of the $s_i=\pm 1$ configurations is enhanced
by making these states $r$-fold degenerate. Schupper and Shnerb chose
the interactions $J_{ij}$ from a fully-connected Gaussian
distribution, thus obtaining a {\em spin-glass}-like
phase~\cite{CrLe}.  If instead one wishes to model a {\em structural}
(fragile) glass behavior, one may choose interactions as in either the
random orthogonal model~\cite{MaPaRiI}, or to consider a $p$-spin
interaction model with spin-1 variables like in Ref.~\cite{Mo}.  We
have studied the equilibrium phase diagram of the former model in
detail and found, for large enough $r$, a reentrant behaviour in both
the dynamic and static glass transition line. The structural glass
transition in this model can be either thermodynamically first or
second order (i.e.  with or without latent heat), depending on the
value of $D$~\cite{Seku}.

\begin{figure}
\begin{center}
\includegraphics[width=9.cm]{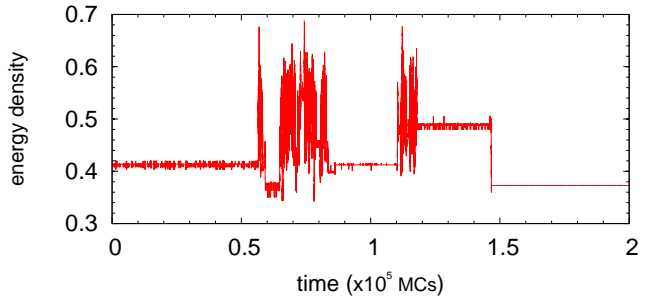}
\end{center}
\caption{
  Energy density {\em vs.} time after a system (size $N=128$) in
  the liquid phase at $T=0.03$ and $D=3.0$ was taken to the jammed
  phase by a stirring force with $\epsilon=1.6$ and $\delta=0.2$.}
\label{uno}
\end{figure}

{\bf Shear-thickening models.} As mentioned above, stirring is
somewhat analogous (and in the example above exactly equivalent) to
coupling to a high temperature bath.  One can thus imagine that in a
problem with phase reentrance, stirring might induce a transition from
the liquid to the glassy phase.  This is clear in the folded polymer
problem described above: taking into account the known fact that
shearing~\cite{deGennes}, or random stirring~\cite{tano} can make
polymers unfold to the interacting state, this may result in an
increase in the viscosity of the polymer melt.  Shear-thickening is in
such cases a form of phase reentrance~\cite{Ca}.

What we have discussed so far suggests that one can model
shear-thickening by considering a nonconservative forcing acting on a
reentrant model, for example a force field $f_i^{\scriptscriptstyle
  \rm stir}$ acting on the $i^{\scriptscriptstyle \rm th}$ spin
of~(\ref{H}):
\begin{equation}
  f_i^{\scriptscriptstyle \rm stir}= \epsilon \sum_j
    J^{\scriptscriptstyle \rm as}_{ij}
    \left(1-s_j^2\right) + \delta \sum_j
    K^{\scriptscriptstyle \rm as}_{ij} s_j \,,
  \label{sti}
\end{equation}
where $J^{\scriptscriptstyle \rm as}_{ij}=-J^{\scriptscriptstyle \rm
  as}_{ji}$ and $K^{\scriptscriptstyle \rm
  as}_{ij}=-K^{\scriptscriptstyle \rm as}_{ji}$ are independent
Gaussian random variables with zero mean and variance $1/N$.  The two
components of the force field act independently on the `folded' and
the `unfolded' configurations with stirring strengths $\epsilon$ and
$\delta$ respectively~\cite{note2}.

{\bf Phase diagram - metastability and coexistence.}  Fig.~\ref{uno}
shows the evolution of the energy of a small system after stirring
terms (\ref{sti}) were applied to the noninteracting `liquid' state.
The fraction of $\pm 1$ spin becomes appreciable and in the energy
{\em vs.}~time plots we observe intermittent arrest and flow behaviour
with jumps between long-lived interacting states.  If the temperature
is low and stirring is not too strong, the system quickly falls in a
state that is for all practical purposes stable.  Increasing the
stirring strength the trapping times become shorter, and for
sufficiently high stirring rates the system becomes a normal
(non-aging) liquid. We shall not discuss in detail the shear-thinning
(or rejuvenation) aspect, as it has been already extensively discussed
in the literature~\cite{Babeku,Babe}.  Intermittent situations where
the system jams and unjams as in Fig.~\ref{uno} have been
observed~\cite{frith}.

\begin{figure}
\begin{center}
\includegraphics[width=8.5cm]{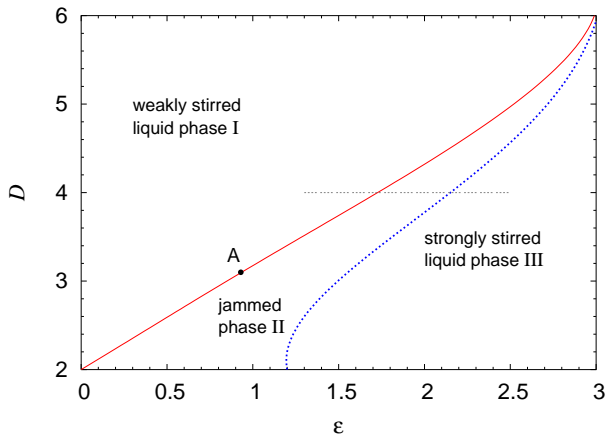}
\end{center}
\caption{A section of the dynamical phase diagram for
  temperature $T=0.05$, $\delta=0.0$ and $r=6$. The transition between
  the phases I and II is first-order below the point A and
  second-order above it.}
\label{dos}
\end{figure}

The analysis of the temporal evolution of the energy and the spin-spin
correlation function allows to identify different dynamical regimes
and construct the dynamical phase diagram by varying the several
parameters characterizing the system.  Fig.~\ref{dos} shows a section
of such a dynamical phase diagram in terms of $D$ and stirring
strength $\epsilon$.  The weakly stirred liquid phase I has a low
density of interactive sites $\rho \simeq 0$. The `jammed' phase II is
characterized by a fraction of spins in the interactive $\pm 1$ state
and aging (i.e. the progressive trapping in ever deeper interacting
states), and the `liquid' III regime for high stirring rates is just
the result of shear-thinning of the jammed state.
For $D<2$ and low temperature the system is glassy in the absence of
stirring. On increasing $D$ and the stirring force there is `first
order' jamming transition (below the point {\small A} in
Fig.~\ref{dos}) with hysteresis in $\epsilon$: along this curve the
liquid and jammed phases coexist.  Above the point {\small A} the
transition from phase I to phase II is a `second order' jamming
transition without hysteresis, but with a regime in which the system
forms under stress an aging glass.  For much larger values of $D$
there is no jamming for any stirring, but there is continuous
shear-thickening when the $T-\epsilon$ trajectory followed passes near
a transition line in phase space.
 
\begin{figure}
\begin{center}
\includegraphics[width=8.5cm]{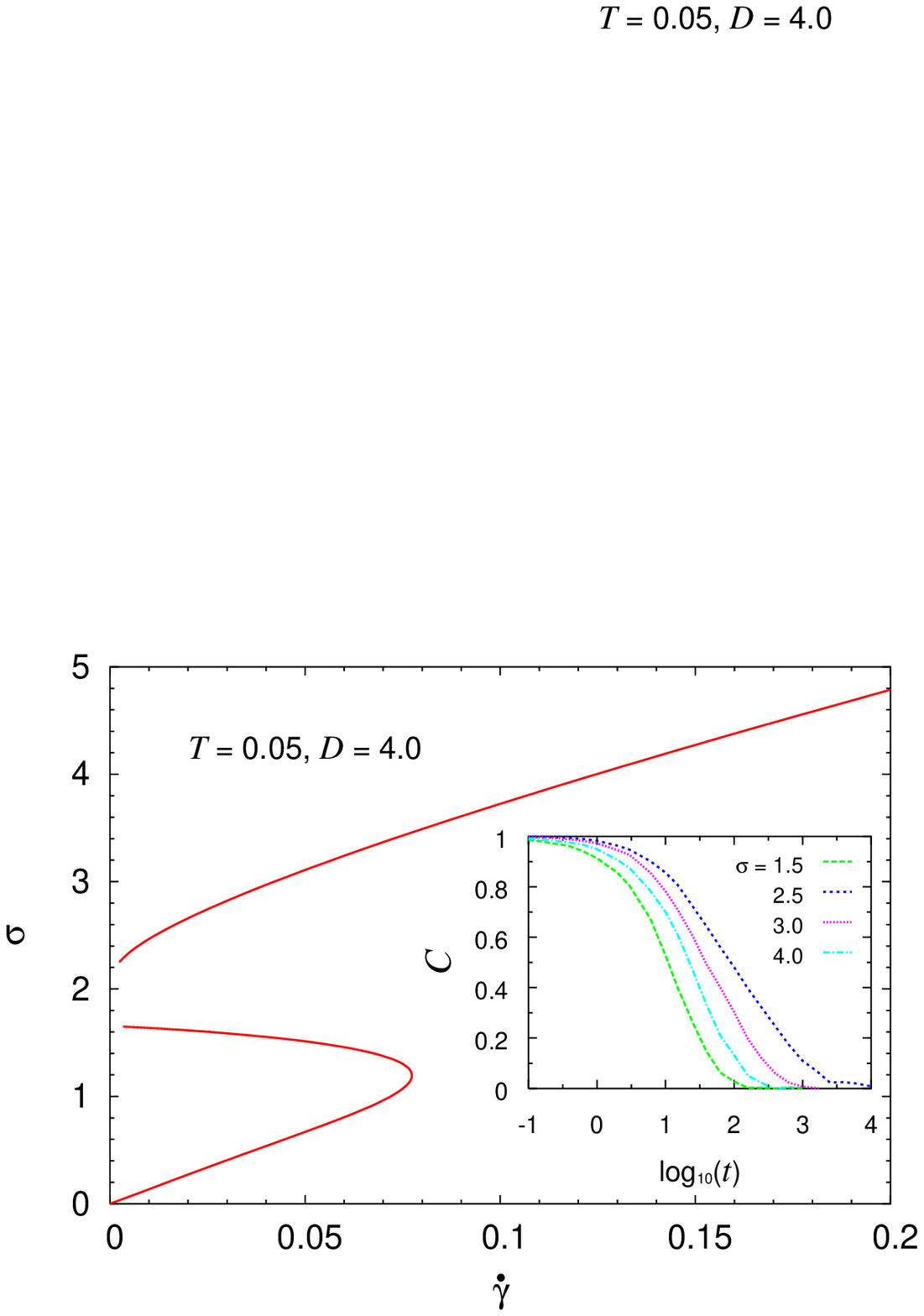}
\end{center}
\caption{Flow curve along the constant $D=4$ line for 
  $T=.05$.  $\sigma \equiv \epsilon$ and $\dot{\gamma} \equiv
  \sigma/\tau_{\alpha}$ are the analogue of stress and shear rate for
  a mean-field system. Note how the intersection of the line $D=4$
  with the jammed phase in Fig.~\ref{dos} is reflected here. The inset
  shows the corresponding spin-spin correlation function, $C$, vs time
  $t$, for increasing stress (system size $N=500$).}
\label{doss}
\end{figure}

In these mean-field models the dissipated power scales as $
\epsilon^2/\tau_{\alpha}$, where $\tau_{\alpha}$ is the
$\alpha$-relaxation time of the system~\cite{Babeku}.  Comparing to a
standard shear flow this suggests that the amplitude of the driving
force, $\epsilon$, plays the role of a stress, $\sigma$, while
$\epsilon/\tau_{\alpha}$ is analogous to a shear rate, $\dot{\gamma}$.
We may thus obtain the standard $\sigma$ vs.  ${\dot{\gamma}}$ flow
curves by increasing the stirring rate at constant $D$: they turn out
to be strikingly similar to those of Ref.~\cite{holmes}.  In the main
frame of Fig.~\ref{doss} we show an example of such flow curves
corresponding to the 'full jamming' scenario of Ref.~\cite{Armand}. In
this case, one observes an interval of stress, in Fig.~\ref{doss}
between 1.65 and 2.2, within which the flow rate vanishes, even if the
system is ergodic at rest.  The relaxation time, $\tau_{\alpha}$, was
estimated as the time integral of the normalized spin-spin correlation
function.  Examples of correlation curves are shown in the inset of
Fig.~\ref{doss}: Shear-thickening behaviour (a slower decay of $C$) is
observed when $\sigma$ increases from 1.5 to 2.5, while shear-thinning
(a faster decay of $C$) appears for higher stress (for $\sigma$
increasing from 2.5 to 4 in the inset of Fig.~\ref{doss}).

In Ref.~\cite{bibette}, an experiment is described in which a
concentrated suspension of non-Brownian particles is driven by
stirring from the liquid to a metastable jammed phase.  It would be
interesting to see experimentally whether the opposite situation, when
the liquid is the metastable phase, may occur.

\begin{figure}
\begin{center}
\includegraphics[width=8.5cm]{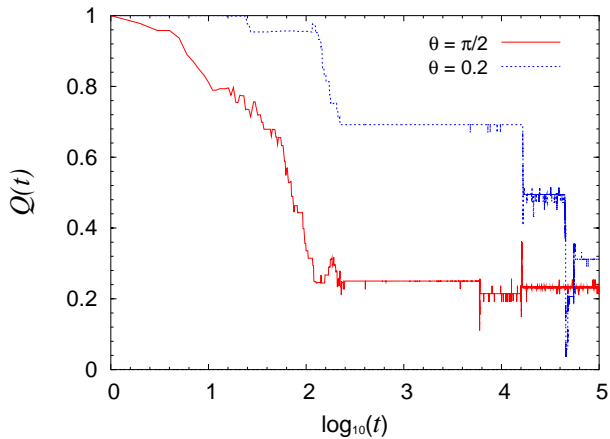}
\end{center}
\caption{Chain fragility: mutual correlation, $Q(t)$, of two identical
  jammed systems, in one of which the stirring direction has been
  changed at time $t=0$. $\theta=\pi/2$ corresponds to the case in
  which the two systems have independent random stirring forces, while
  in the case $\theta=0.2$ the random stirring forces are correlated.}
\label{tres}
\end{figure}

{\bf Chain fragility and aging of the jammed phase.} One of the
properties of materials that are jammed by stirring that we may wish
to test in this model is the fragility with respect to incremental
stresses in a different direction~\cite{boca}.  If in a system in
phase II we change the realization of stirring forces from
$J^{\scriptscriptstyle \rm as}_{ij}$ to $J^{\scriptscriptstyle \rm
  as}_{ij} \cos \theta + J'^{\scriptscriptstyle \rm as}_{ij} \sin
\theta $, and similarly for $K^{\scriptscriptstyle \rm as}_{ij}$, even
for small $\theta$ we find that the system responds by rearranging its
configuration, the faster the larger the value of $\theta$ ($\theta
\in [0,\pi/2]$), see Fig.~\ref{tres}.

{\bf Conclusions.} In this Letter we discussed a connection between
the mechanisms of entropy-driven phase reentrance and
shear-thickening.  This relation may exist in some cases just in
principle, as the temperatures or chemical potentials needed to
actually affect substantially the particles may be in practical
situations extremely high.  The present models are clearly schematic,
but not much has been put into them and yet we see the elementary
constituents self-organize to produce stirring-induced jamming with
aging and intermittency, non-Newtonian rheological behaviour like
shear thinning and thickening, metastability and chain fragility.

Let us finally mention that one can also construct a reentrant
continuous model with $p$-spin interactions $H_p$ following the same
idea outlined above \cite{BeKuSe}.  From this model one can
immediately obtain Mode-Coupling with reentrance --- and also shear
thickening by adding stirring forces of the form~(\ref{sti}) to the
Langevin dynamics $ \dot x_i= -\partial_{x_i} H_p + \eta_i $.  Another
interesting approach is to introduce stirring terms in the
Mode-Coupling models of colloids with short-range attractive
potentials~\cite{fuchs}.  This would allow to investigate the
intriguing perspective that stirring can drive the liquid-liquid, the
liquid-glass and also the glass-glass transition in such systems. Work
along these lines is in progress \cite{BeKuSe}.

\medskip

We thank A. Ajdari, D. Bonn, B. Cabane, M.~Cates and E.~Zaccarelli for
discussions and suggestions.  MS gratefully acknowledges support of
the {\small EVERGROW} project and hospitality at {\small ESPCI}. JK
acknowledges the hospitality of the Granular Matter program at {\small
  KITP}, University of California at Santa Barbara.


\begin{thebibliography}{40}
\bibitem{note2} If instead the $s_i$ are interpreted as representing a
  particle when $s_i=\pm 1$ and a hole if $s_i=0$, this stirring force
  can be viewed as a coupling to a particle reservoir.
\end{thebibliography}
\end{document}